\documentclass[reprint,
superscriptaddress,
 amsmath,amssymb,
 aps,
 prl,
]{revtex4-2}
\usepackage{hyperref}
\usepackage{graphicx}
\usepackage{dcolumn}
\usepackage{bm}
\usepackage{amsmath} 

\begin{document}


\title{Transition from Collective to Local Radial Motional Modes in a Tapered Paul Trap}

\author{Manika Bhardwaj}
\thanks{These authors contributed equally}
\affiliation {Experimental Physics I, University of Kassel, Heinrich-Plett-Strasse 40, 34132 Kassel, Germany}
\author{Moritz G{\"o}b}
\thanks{These authors contributed equally}
\affiliation {Experimental Physics I, University of Kassel, Heinrich-Plett-Strasse 40, 34132 Kassel, Germany}
\author{Bogomila S. Nikolova}
\affiliation{Center for Quantum Technologies, Department of Physics, St. Kliment Ohridski University of Sofia, James Bourchier 5 blvd, 1164 Sofia, Bulgaria}
\author{Bernd Bauerhenne}
\affiliation {Experimental Physics I, University of Kassel, Heinrich-Plett-Strasse 40, 34132 Kassel, Germany}
\author{Peter A. Ivanov}
\affiliation{Center for Quantum Technologies, Department of Physics, St. Kliment Ohridski University of Sofia, James Bourchier 5 blvd, 1164 Sofia, Bulgaria}
\author{Kilian Singer}
\email[E-Mail: ]{ks@uni-kassel.de}
\affiliation{Experimental Physics I, University of Kassel, Heinrich-Plett-Strasse 40, 34132 Kassel, Germany}

\date{\today}

\begin{abstract}
With coupled detuned oscillators, either individual or collective oscillations are observable. The latter is used in quantum information processing in linear Paul traps. Here, we study the transition from collective radial modes at stronger axial confinements into individual radial oscillations at low axial confinements in a tapered Paul trap. The eigenmodes are experimentally studied in detail in the transition regime and compared with theoretical predictions. The features studied will enable investigation of modified heat transport phenomena and defect formation in trapped ions.
\end{abstract}

\maketitle
Trapped ions are a formidable model system for the study of quantum effects that can be efficiently manipulated by either laser light or microwave \cite{Bruzewicz2019}. Mutual coupling between the ions is realized through the use of quantized motional modes in either axial or radial directions, allowing for efficient implementation of high fidelity multi qubit gates. Such gates routinely reach errors below $10^{-3}$ in state-of-the-art experiments \cite{Ballance2016, Gaebler2016, Clark2021, loeschnauer2026}, making trapped ions one of the leading platforms for scalable quantum information processing. The use of radial modes and spin-dependent squeezing operations has the potential to realize fast quantum gates \cite{Zhu2006,Haffner2008} and programmable $N$-body interaction in a robust fashion \cite{Katz2023}. A tapered ion trap \cite{Deng2024} is of particular interest, as it features an axial position dependent radial confinement, for which advanced protocols can be devised. This trap design enabled us to realize a single atom heat engine \cite{Rossnagel2016} and amplification of zeptonewton forces \cite{Deng2023}, and allows for the investigation of thermodynamic cycles, such as the Otto cycle \cite{Abah2012} in the quantum regime \cite{Rossnagel2014,Kosloff2017,Levy2020,Izadyari22}. Further, this geometry is of interest for the study of heat transport beyond measurement schemes proposed for linear traps \cite{Bermudez2013} as it might exhibit asymmetric heat transport \cite{Simon2019}. The new quantum mechanical features of a tapered ion trap are theoretically studied in \cite{Nikolova2025}, predicting a mechanically squeezed Kerr oscillator, where the coupling between the axial and radial degrees of freedom leads to a Kerr nonlinearity of the radial mode with magnitude controlled by the trapping frequency. This allows for the realization of non-Gaussian quantum gates with continuous variables \cite{Braunstein2005}. 

Here, we theoretically model the radial eigenfrequencies and eigenvectors of ions in a tapered Paul trap and compare them with experimental results of three $^{40}$Ca$^+$ ions. 
The radial eigenmodes are characterized across coupling regimes ranging from strong axial confinements, exhibiting collective radial modes, to weak axial confinements, featuring local radial excitations. We observe a complete transformation of eigenfrequencies and eigenvectors across these regimes, consistent with theoretical predictions, laying the technological foundation for future quantum experiments to use these features.
Our characterization of the collective-to-local radial eigenmodes thus directly supports the controlled use of those Kerr-nonlinear radial modes as resources for continuous-variable quantum computing with trapped ions \cite{Menicucci2014}.

\begin{figure}[h]
    \centering
    \includegraphics[scale=1]{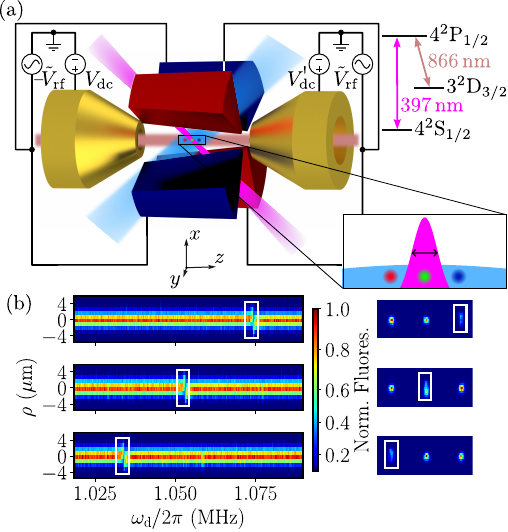}
    \caption{(a) Schematic depiction of the experiment. The tapered ion trap consists of two pairs of rf electrodes, here shown in red and blue, supplied with an rf voltage with a relative phase of $\pi$. Axially, the ions are confined by dc voltages to the golden endcap electrodes. The spatially broad beam depicted in light blue is used for imaging and excitation of the motion of all three calcium ions. The additional magenta narrow beam depicts the focused beam used to excite the middle ion. Both lasers are close to the Doppler transition of $^{40}$Ca$^+$ at $397$\,nm. Additionally a repumper at $866$\,nm is aligned through the hollow center of the endcap electrodes, here depicted in light red. The inset illustrates the intensity profiles of the two Doppler-cooling lasers (blue and magenta). Furthermore, it depicts the three ions, with the blue, green, and red denoting the ion closer to the apex, in the middle, and towards the open side of the trap, respectively. The color scheme denoting the three ions will be used throughout this paper. (b) Local excitations of the three ions at lowest axial confinement using the broad beam. On the left, the radial projection at each driving frequency $\omega_d$ is shown for the three ions. The white boxes highlight the resonance of each ion. On the right, the corresponding camera images are depicted at the respective resonance frequencies.}
    \label{fig:schematic}
\end{figure}

The tapered Paul trap used in our experiment is depicted in Fig.\,\ref{fig:schematic}(a). It consists of four radio-frequency (rf) electrodes, which are aligned at an angle of $10^\circ$ to the trap axis, and two endcap electrodes. Each pair of diagonally opposite rf electrodes is supplied with the rf voltage $\tilde{V}_\text{rf}$ with a phase of $\pi$ between the two pairs. This is done in order to minimize the micromotion along the axial direction \cite{Deng2024}. The distance between the electrodes of a pair at the narrowest point is $1$\,mm. The axial confinement is generated by the direct current (dc) voltages supplied to the endcap electrodes, which are spaced $4.8\,\mathrm{mm}$ apart. 
The geometry of the trap leads to a funnel-shaped potential. The effective trapping potential is
\begin{equation}
    V_\mathrm{tr}=\frac{m}{2}\left[\left( 1+2\frac{z}{\ell_0}\right)\left(\omega_x^2x^2+\omega_y^2y^2\right) +\omega_z^2z^2\right],
\end{equation}
where $m$ is the mass of the ion, $\ell_0=1.81\,$mm \cite{Deng2023} denotes the funnel-length, $\omega_{x}$ and $\omega_y$ are the trapping frequencies in radial directions and $\omega_z$ is the trapping frequency in the axial direction. 
In our setup with $^{40}$Ca$^+$ ions, the radial and axial trapping frequencies are typically in the order of  $\simeq 2\pi\times 1$\,MHz and $ \simeq 2\pi\times 100$\,kHz, respectively. 

A Doppler beam at $397$\,nm, red-detuned from the transition $4^{2}\text{S}_{1/2}\leftrightarrow4^{2}\text{P}_{1/2}$ by twice its natural linewidth, is aligned at $45^\circ$ to the axial direction of the trap, as illustrated in Fig.\,\ref{fig:schematic}(a). Its broad focus is used to address all three ions, and it is henceforth referred to as the broad beam. An additional tightly focused $397$\,nm beam with a radius of $17.0(2)\,\mu$m is oriented perpendicular to the broad beam, enabling single-ion addressing. This beam is further red-detuned with respect to the broad beam by $\sim 6$\,MHz. The repumper beam at $866$\,nm is aligned via the through-hole in the endcap electrodes.

Contrary to previous work, in which single $^{40}$Ca$^+$ ions were used in experiments \cite{Deng2024, Rossnagel2016, Deng2023}, we study the dynamics of multiple ions here. Since each ion carries a charge $e$, the Coulomb interaction between the $N$ ions must be included in the potential, which reads as:
\begin{align}
    V=\sum_i^N&\left[ V_\mathrm{tr}(\vec{r}_i)
     + \frac{e^2}{8\pi\varepsilon_0} \sum_{j\neq i}^N\frac{1}{\left|\vec{r}_i-\vec{r}_j\right|}\right],
     \label{eq:TotalPot}
\end{align}
with $\varepsilon_0$ denoting the vacuum permittivity.
As the confinement along the axial direction is lower than in the radial directions, the ions are aligned in a one-dimensional chain along the axis. 
Therefore, the position of the $i$-th ion is expressed as $\vec{r}_i= \tilde{x}_i\hat{e}_x+\tilde{y}_i\hat{e}_y+(z_i^0+\tilde{z}_i)\hat{e}_z$, where $z_i^0$ denotes the equilibrium position and $\{\tilde{x}_i,\tilde{y}_i,\tilde{z}_i\}$ the displacement from it. The equilibrium positions can be found by determining the minima of Eq. (\ref{eq:TotalPot}) \cite{James1998}. 
In order to describe the coupling between the ions, the equations of motion are Taylor-approximated around the equilibrium positions, and the Hessian matrices $A_{i,j}^{(q)}=\partial^2V/\partial q_i\partial q_j$ with $q=\{x,y,z\}$  are computed. In the axial direction, radial terms vanish in second order, leading to a Hessian matrix resembling that of a linear ion trap \cite{Nikolova2025}. However, in the radial directions $\rho = \{x,y\}$, due to the tapered geometry of the trap, we obtain:
\begin{equation}
A_{i,j}^{(\rho)}=\left\{
\begin{array}{rl}
    1+\frac{2 \lambda u_{i}}{\ell_{0}}-\displaystyle{\sum_{k\neq i}^{N}}\frac{\beta^{2}_{\rho}}{|u_{i}-u_{k}|^{3}},\ &\mathrm{for}\ (i=j) \\
\displaystyle\frac{\beta^{2}_{\rho}}{|u_{i}-u_{j}|^{3}},\ &\mathrm{for}\ (i\neq j).
\end{array}
\right.
\label{eq:matrix}
\end{equation}
Here we are using, $\beta_\rho=\omega_z/\omega_\rho$, the length scale $\lambda^3=e^2/4\pi\varepsilon_0m\omega_z^2$ and the dimensionless equilibrium position $\lambda u_i=z_i^0$.
The eigenfrequency of the $i$-th eigenmode $\Omega_{\rho,i}=\sqrt{\gamma_{\rho,i}}\,\omega_\rho$, where $\gamma_{\rho,i}$ is the $i$-th eigenvalue of $A^{(\rho)}$.
The normal mode eigenvectors $A^{(\rho)}\vec{a}_i=\gamma_{\rho,i}\vec{a}_i$ describe the participation of each ion in the $i$-th eigenmode. Increasing the axial confinement not only lowers the spacing between the ions (thereby increasing their coupling), but also weakens the effective radial confinement $\omega_\rho=(\omega_{\rho,0}^2-\omega_z^2/2 )^{1/2} $ \cite{Drewsen2000, Gulde2003}.

\begin{figure}[ht]
    \centering
    \includegraphics[scale=1]{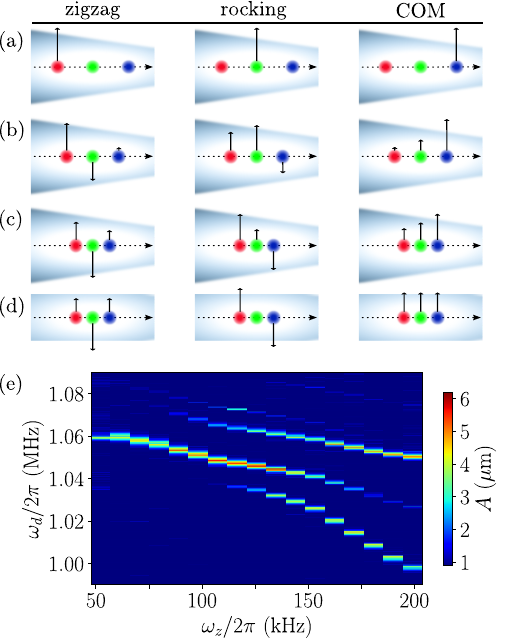}
    \caption{Illustration of radial eigenmode oscillations of a tapered trap at axial confinements of $40$\,kHz (a), $125$\,kHz (b), and 200\,kHz (c), compared to the radial eigenmode oscillations of a linear trap (d). (e) Plot of the sum of the oscillation amplitudes of the three ions at each $\omega_d$ over the measured range of $\omega_z$, while the middle ion is driven with the intensity-modulated focused beam. The outer ions couple to the middle ion with increasing $\omega_z$, exciting oscillations at different eigenfrequencies.
    }
    \label{fig:ExplenationEigenvectors}
\end{figure}

At low axial confinements, the ions are far apart. 
Due to the funnel-shaped potential, the ions experience different radial confinements at their equilibrium positions. 
The Coulomb interaction is comparably small to this energy splitting, resulting in localized oscillations, which are depicted in Fig.\,\ref{fig:schematic}(b) and Fig.\,\ref{fig:ExplenationEigenvectors}(a). Increasing the axial confinement reduces the distance between the ions, which increases their Coulomb interaction. This leads to an onset of collective oscillations, depicted in Fig.\,\ref{fig:ExplenationEigenvectors}(b). The higher the confinement gets, the stronger the ions start to oscillate collectively (Fig.\,\ref{fig:ExplenationEigenvectors}(c)), similarly to a linear ion trap (Fig.\,\ref{fig:ExplenationEigenvectors}(d)). The transition from individual to collective oscillations can also be described by the dynamics of local radial phonons \cite{Porras2004, Ivanov2009}. Indeed, at low axial frequencies, a large frequency gap exists between the individual position-dependent phonon frequencies, which suppresses phonon hopping between ions and leads to phonon localization. Increasing $\omega_{z}$, the phonon frequency gap decreases, resulting in delocalization of the local phonons and collective oscillation behavior of the ion chain. \\

In the following, we describe our measurement protocol for determining the radial eigenmodes of three $^{40}$Ca$^+$ ions across different axial confinements in the tapered trap, ranging from 47\,kHz to 205\,kHz. Both endcap voltages of the trap are adjusted individually to vary the axial confinement in such a way that the position of the middle ion is not shifted. 

At each axial confinement, the axial center-of-mass (COM) mode is measured by sinusoidally modulating the intensity of the broad $397$\,nm beam. At different intensities, the number of photons absorbed by the ions varies, which in turn changes the number of momentum kicks the ions experience. This constitutes an effective driving force with frequency $\omega_{d}$, being scanned in the range of the axial trapping frequency $\omega_z$. At each driving frequency, an image with integration time of $0.2\,$s is taken by an intensified charge-coupled-device (iCCD) camera. When the driving force is resonant with the axial COM mode, the ions oscillate with the highest amplitude. 

The radial eigenmodes are measured using the same method as described above for different axial confinements. One of the two radial principal axes is aligned parallel to the imaging plane to maximize the oscillation amplitude along that radial axis. This is useful in resolving even small oscillation amplitudes. One measurement at the lowest axial confinement is depicted in Fig.\,\ref{fig:schematic}(b). The broad beam is used to excite the oscillations at axial confinements lower than $135$\,kHz, as the driving force is exerted on all ions. However, at higher axial confinements, the out-of-phase oscillations cannot be excited with the global drive anymore.
To excite these out-of-phase oscillations, local addressing is required. This is done by modulating the intensity of the focused $397$\,nm beam, centered on the middle ion. To also cool and image the outer two ions, the broad beam is kept at a constant low intensity while the focused beam is modulated. The sum of the oscillation amplitude of all three ions excited by the focused beam is depicted at each $\omega_d$ in Fig.\,\ref{fig:ExplenationEigenvectors}(e) for different axial confinements.

To determine the eigenfrequencies $\Omega_\rho$ and eigenvector components of the radial modes at each axial confinement $\omega_z$ from the recorded images, further evaluation is needed. At each driving frequency $\omega_d$, the recorded image is summed row-wise in the axial direction for each ion to obtain the radial fluorescence distribution. This is then fitted with a convolution of a Gaussian point spread function and a position-dependent intensity modulation function. This way, the oscillation amplitude $A$ in the radial direction of each ion at each $\omega_d$ is obtained. 

Next, the oscillation amplitude of all ions is added up at each $\omega_d$ and fitted by the sum of three Lorentzian functions with an added constant offset. This way, the radial eigenfrequencies $\Omega_\rho$ of the three ions, which are depicted by the squares and circles in Fig.\,\ref{fig:eigenfrequencies}, are obtained.
More details can be found in the supplemental material \cite{Sup}. The theoretical curves are also displayed in this figure as the blue, green, and red solid lines, representing the eigenmode with the highest, middle, and lowest frequency, respectively, which was fitted to the experimental values by adjusting $\omega_{\rho,0}=1.057$\,MHz. The gray dashed lines represent the eigenfrequencies of a linear trap with a radial confinement of $\omega_{\rho,0}$. The taper is disabled in this case by letting $\ell_0\rightarrow\infty$ in Eq.\,\ref{eq:matrix}. The radial eigenfrequencies of the tapered trap are diverging at low axial confinements as compared to the radial eigenfrequencies of a linear trap, which are converging. As the axial confinement of the tapered trap is increased, the radial eigenfrequencies start approaching those of a linear trap, as can be seen in Fig.\,\ref{fig:eigenfrequencies}.

\begin{figure}[ht]
    \centering
    \includegraphics[width=1\linewidth]{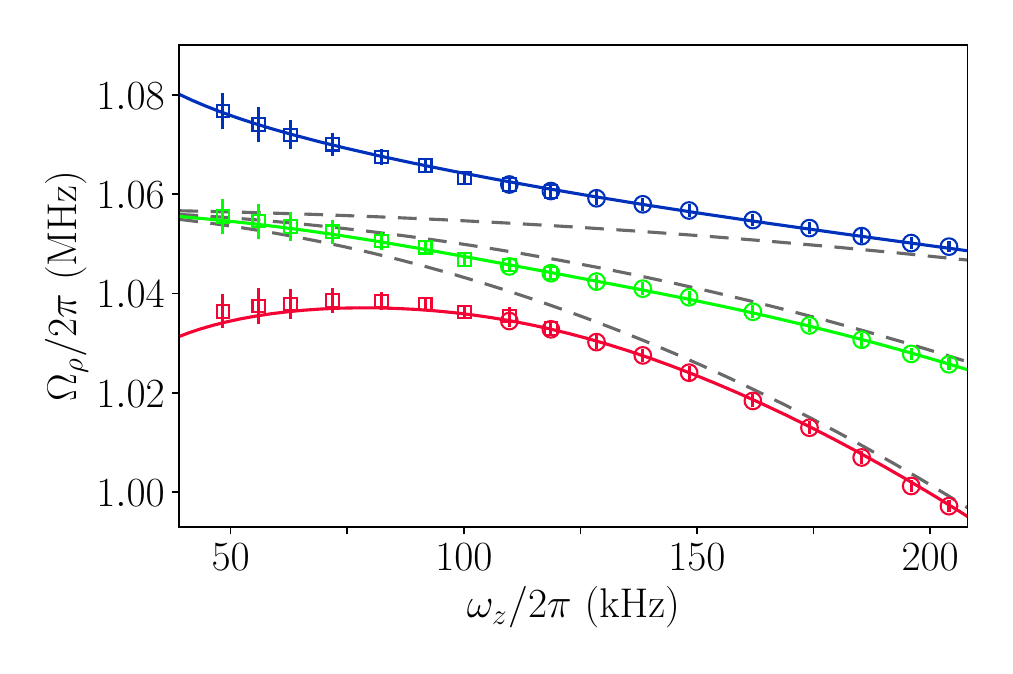}
    \caption{Radial eigenmode frequencies $\Omega_\rho$ of the three ions at different axial confinements $\omega_z$. The squares and circles represent the broad and focused beam measurements respectively, with the error bars denoting the standard deviation. The upper blue (middle green, lower red) solid lines depict the theoretical predictions of the highest (middle, lowest) eigenfrequency of the three ion crystal. At low axial confinements, the colors correspond to the local oscillations of each ion as depicted  in Fig.\,\ref{fig:schematic}. The gray dashed lines illustrate the eigenmodes for a linear Paul trap.
    }
    \label{fig:eigenfrequencies}
\end{figure}

\begin{figure}[ht]
    \centering
    \includegraphics[width=\linewidth]{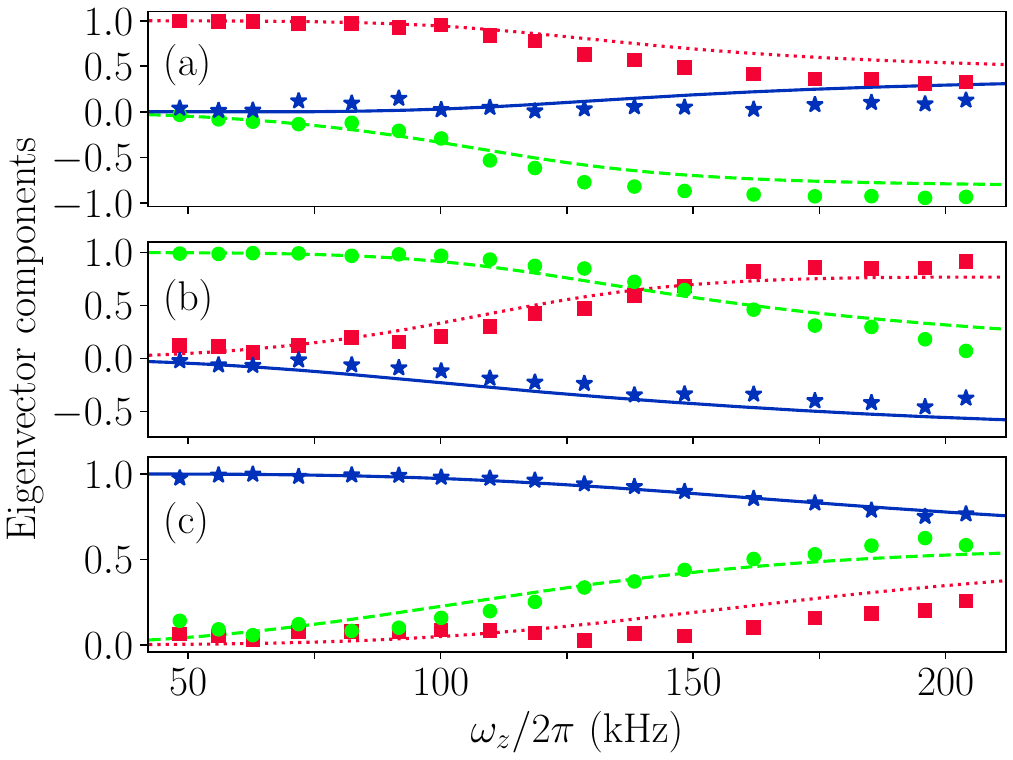}
    \caption{
    The eigenvector components of the three ions at different axial confinements $\omega_z$. Blue stars, green dots and red squares are the experimental values of the eigenvector components of the ion on the apex of the taper, middle ion and the ion on the open side of the taper, respectively. The blue-solid, green-dashed and red-dotted lines are the theoretical values of the eigenvector components, corresponding to the respective ions (Fig.\,\ref{fig:schematic}). At higher axial confinements the zigzag (a), the rocking (b) and the COM radial modes (c) are observed.
    }
    \label{fig:eigenvector entries}
\end{figure}

To gain further insight, we also determine the eigenvector components, which are depicted in Fig.\ref{fig:eigenvector entries}. Now, the oscillation amplitude of each ion is fitted with the Lorentzian curves described above for the whole range of $\omega_d$. However, $\Omega_{\rho}$ is kept constant here, with only the height and half-width at half-maximum varied. The height obtained from the Lorentzian fit for each eigenmode can be interpreted as the ion's maximum oscillation amplitude. These maximum oscillation amplitudes are rescaled such that the corresponding eigenvectors are normalized. The signs are inferred from the relative phase between the ions' oscillations. Further details are presented in the supplemental material \cite{Sup}. The blue stars, green dots, and red squares denote the participation of the right, middle, and left ion for each radial mode in Fig.\ref{fig:eigenvector entries}. The theoretical values are displayed by the blue solid, green dashed, and red dotted lines, respectively.

At low axial confinements, only one ion participates in each eigenmode, exhibiting local oscillations. As the axial frequency is increased, the coupling between the ions increases, leading to oscillations of more than one ion in each radial eigenmode. When the confinement is further increased, all ions begin participating in each radial eigenmode, exhibiting collective oscillations. This differs from a linear trap, where the radial trapping potential is the same for all ions. As a result, the ions exhibit collective motion in the radial direction regardless of the axial confinement. It is also interesting to note that in the COM mode in Fig.\,\ref{fig:eigenvector entries}(c), the ion (blue) at the highest radial confinement is oscillating with the largest amplitude. This behavior is in agreement with the theoretical model presented in this work. Deviations from the theoretical predictions in Fig.\,\ref{fig:eigenvector entries} are due to systematic errors, such as excessive damping caused by the broad imaging beam during the focused beam measurement. Further, partial illumination of the outer two ions by the focused beam at high axial confinements causes an overall lower oscillation amplitude for out-of-phase motions. One way to address this issue is to use a beam with a smaller waist, so that the outer two ions are not illuminated.  

In this paper, we have presented, both theoretically and experimentally, the radial eigenfrequencies and eigenvector components of three $^{40}\text{Ca}^{+}$ ions in a tapered Paul trap. The special geometry of our tapered trap enables a smooth transition of radial modes from the collective to the local regime by varying the axial confinement. The experimental results are well confirmed by the theoretical model. The effect of the taper at weak axial confinements leads to a local change of the radial eigenfrequencies along the trap axis, resulting in the observation of local radial modes with low cross-coupling to neighboring ions. As the axial confinement is increased, the radial modes approach the collective regime, as observed in a linear Paul trap. Global and selective ion addressing were necessary to study the eigenvectors, which shed light on the markedly different behavior of local vs. collective modes with respect to the dynamics of individual ions. Experimental and theoretical details are provided to enable the exploitation of these results to enhance current quantum information processing technologies. For example, control over the radial motional modes can be used to suppress unwanted crosstalk during state preparation. Multi-frequency excitation of the radial modes might enable engineering of specific spin-spin couplings beyond nearest neighbors. Transition to individual ion oscillation is suitable for faster cooling since only one ion participates in the radial motion. Also of interest is the study of heat transport with individual ions across different coupling regimes \cite{Bermudez2013, Simon2019}, potentially including structural phase transitions \cite{Ruiz2014}. Furthermore, topological defects, as described by the Kibble-Zurek scaling \cite{Pyka2013, Ulm2013}, might be affected during the transition from linear to tapered geometry, which could provide interesting insights into the mechanisms involved, given the additional symmetry breaking due to the taper. The presented models could also inspire novel trapping geometries that allow changing the funnel length $\ell_0$ dynamically, thereby forcing a transition between collective and local regimes without changing the ion distance. As the transition regime before entering the collective oscillator regime can be described by weakly coupled local oscillators with slightly detuned frequencies, the study of synchronization \cite{Pikovsky2001} in a tapered trap might become feasible. Extension into the quantum regime could lead to an amplification of synchronization phenomena \cite{Solanki2022,Aifer2024,Li2025}.

\section*{Acknowledgments}

We acknowledge funding by the Deutsche Forschungsgemeinschaft (DFG, German Research Foundation)—Projects No. 499241080 and No. 384846402—through the QuantERA grant ExTRaQT and the Research Unit Thermal Machines in the Quantum World (FOR 2724). 
B. S. N. and P. A. I. acknowledge the Bulgarian national plan for recovery and resilience, contract BG-RRP-2.004-0008-C01 (SUMMIT: Sofia University Marking Momentum for Innovation and Technological Transfer), project number 3.1.4.

\nocite{*}
\bibliography{TaperedChain}

\end{document}